\documentclass[pra,twocolumn,amsmath,amssymb,superscriptaddress,
showpacs]{revtex4}
\usepackage{graphics}
\usepackage{epsfig}
\usepackage{bbm}
\usepackage[latin1]{inputen c} 

\newcommand{\be}{\begin{equation}}
\newcommand{\ee}{\end{equation}}
\newcommand{\eea}{\end{eqnarray}}
\newcommand{\bea}{\begin{eqnarray}}

\newcommand{\va}[1]{\ensuremath{(\Delta#1)^2}}

\newcommand{\ex}[1]{\ensuremath{\left\langle{#1}\right\rangle}}
\newcommand{\exs}[1]{\ensuremath{\langle{#1}\rangle}}
\newcommand{\mean}[1]{\ensuremath{\langle{#1}\rangle}}

\newcommand{\eins}{\openone}
\newcommand{\qed}{\ensuremath{\hfill \Box}}

\newcommand{\ketbra}[1]{\ensuremath{| #1 \rangle \langle #1 |}}

\newcommand{\ket}[1]{\ensuremath{|#1\rangle}}

\newcommand{\kommentar}[1]{}
\newcommand{\trace}{{\rm Tr}}

\renewcommand{\vr}{\ensuremath{\rho}}

\begin{document}
\title{Spin squeezing and entanglement}
\date{\today}
\begin{abstract}
What is the relation between spin squeezing and entanglement? To
clarify this, we derive the full set of generalized spin squeezing
inequalities for the detection of entanglement. These are
inequalities for the mean values and variances of the collective
angular momentum components $J_k.$ They can be used for the
experimental detection of entanglement in a system of
spin-$\tfrac{1}{2}$ particles in which the spins cannot be
individually addressed. We present various sets of inequalities that
can detect all entangled states that can be detected based on the
knowledge of: (i) the mean values and variances of $J_k$ in three
orthogonal directions, or (ii) the variances of $J_k$ in three
orthogonal directions, or (iii) the mean values of $J_k^2$ in  three
orthogonal directions or (iv) the mean values and variances of $J_k$
in arbitrary directions. We compare our inequalities to known spin
squeezing entanglement criteria and discuss to which extent spin
squeezing is related to entanglement in the reduced two-qubit
states. Finally, we apply our criteria for the detection of entanglement in
spin models, showing that they can be used to detect bound
entanglement in these systems.
\end{abstract}

\author{G\'eza T\'oth}
\email{toth@alumni.nd.edu}
\affiliation{Department of Theoretical Physics, The University of the Basque Country,
P.O. Box 644, E-48080 Bilbao, Spain }
\affiliation{Ikerbasque--Basque Foundation for
Science, Alameda Urquijo 36, E-48011 Bilbao, Spain}
\affiliation{ICFO--The Institute of Photonic Sciences, Mediterranean Technology Park, E-08860
Castelldefels (Barcelona), Spain}
\affiliation{Research Institute for Solid State Physics and Optics,
Hungarian Academy of Sciences, \\ P.O. Box 49, H-1525 Budapest,
Hungary}

\author{Christian Knapp}
\email{christian.knapp@uibk.ac.at}
\affiliation{Institut f\"ur Theoretische Physik, Universit\"at
Innsbruck, Technikerstra{\ss}e 25, A-6020 Innsbruck, Austria,}

\author{Otfried G\"uhne}
\email{otfried.guehne@uibk.ac.at}
\affiliation{Institut f\"ur Theoretische Physik, Universit\"at
Innsbruck, Technikerstra{\ss}e 25, A-6020 Innsbruck, Austria,}
\affiliation{Institut f\"ur
Quantenoptik und Quanteninformation, \"Osterreichische Akademie der
Wissenschaften, A-6020 Innsbruck, Austria}

\author{Hans J. Briegel}
\email{hans.briegel@uibk.ac.at} \affiliation{Institut f\"ur
Theoretische Physik, Universit\"at Innsbruck, Technikerstra{\ss}e
25, A-6020 Innsbruck, Austria,} \affiliation{Institut f\"ur
Quantenoptik und Quanteninformation, \"Osterreichische Akademie der
Wissenschaften, A-6020 Innsbruck, Austria}


\maketitle



\section{Introduction}

Entanglement lies at the heart of many problems in quantum
mechanics and has attracted an increasing attention in recent years
\cite{BE00,entanglement}. Entanglement is needed in several quantum
information processing tasks such as teleportation and certain
quantum cryptographic protocols. It also plays an important role in
quantum computing making it possible that quantum computers can
outperform their classical counterparts for several problems such as
prime factoring or searching. Moreover, entangled states and the
creation of quantum entanglement naturally arise as goals in
nowadays quantum control experiments when studying the non-classical
phenomena in quantum mechanics.

When in an experiment entanglement is created, it is important to
detect it. Thus, in many quantum physics experiments the creation of
an entangled state is followed by measurements. Based on the results
of these measurements, the experimenters conclude that the produced
state was entangled. However, in many-particle experiments the
possibilities for quantum control are very limited. In particular,
the particles cannot be individually addressed. In such systems, the
entanglement can be created and detected with collective operations.

Spin squeezing is one of the most successful approaches for creating
quantum entanglement in such systems
\cite{K93,W94,HS99,VR01,SD01,SM99,SM01,SM01_2,GS04,WS01,KB98,HM04,HP06,EM05,EM08}.
Ref.~\cite{K93} defined spin squeezing in analogy with squeezing in quantum optics:
 Let us consider an ensemble of
$N$ spin-$\tfrac{1}{2}$ particles, and define the observables for
the collective angular momentum as \be J_l:=\frac{1}{2}\sum_{k=1}^N
\sigma_{l}^{(k)} \ee for $l= x,y,z$ and where $\sigma_{l}^{(k)}$ are
Pauli matrices. Then, the variances of the angular momentum
components are bounded by the following uncertainty relation
\begin{equation}
\va{J_z}\va{J_y}\ge \frac{1}{4}\vert\exs{J_x}\vert^2.\label{eqKU}
\end{equation}
If $\va{J_z} := \mean{J_z^2} - \mean{J_z}^2$ is smaller than the
standard quantum limit $\tfrac{1}{2}\vert\exs{J_x}\vert$ then the
state is called spin squeezed \cite{spinsqremark}. In practice this means that the
mean angular momentum of the state is large, and in a direction orthogonal to it
the angular momentum variance is small. An alternative and slightly different
definition of spin squeezing considered the usefulness of spin squeezed states
for reducing spectroscopic noise or to improve the accuracy of atomic
clocks \cite{W94,sdef}.

It has already been noted in Ref.~\cite{K93} that the occurrence of
spin squeezing is connected to the correlations between the spins.
In fact, as shown in  Ref.~\cite{SD01}, there is an entanglement
criterion for the detection of the entanglement of spin squeezed
states: If an $N$-qubit state violates the inequality
\begin{eqnarray}
\frac{\va{J_z}}{\exs{J_x}^2+\exs{J_y}^2}\ge \frac{1}{N},
\label{motherofallspinsqueezinginequalities}
\end{eqnarray}
then the state is entangled (not separable), that is, it can not be
written as \cite{W89} \be \vr = \sum_k p_k \vr_k^{(1)} \otimes
\vr_k^{(2)} \otimes...\otimes \vr_k^{(N)}. \ee where the $p_k$ form
a probability distribution.

After this first entanglement criterion several generalized spin
squeezing criteria for the detection of entanglement appeared in the
literature \cite{GT04,KC05,GT06} and have been used experimentally
\cite{spexp,spexp2}. In Ref.~\cite{GT04}, a generalized spin
squeezing inequality was presented that detects entanglement close
to many-body spin singlets, such as for example, the ground state of
an anti-ferromagnetic Heisenberg chain. In Refs.~\cite{KC05}, a
generalized spin squeezing criterion was presented detecting the
presence of two-qubit entanglement. For symmetric systems, these
criteria are necessary and sufficient. In Ref.~\cite{GT06}, other
criteria can be found that detect entanglement close to symmetric
Dicke states. All these entanglement conditions were obtained using
very different approaches. Therefore, one may ask: Is there a
systematic way of finding all such inequalities? Clearly, finding
such optimal entanglement conditions is a hard task since one can
expect that they contain complicated nonlinearities.

\begin{figure}
\centerline{ \epsfxsize3.3in \epsffile{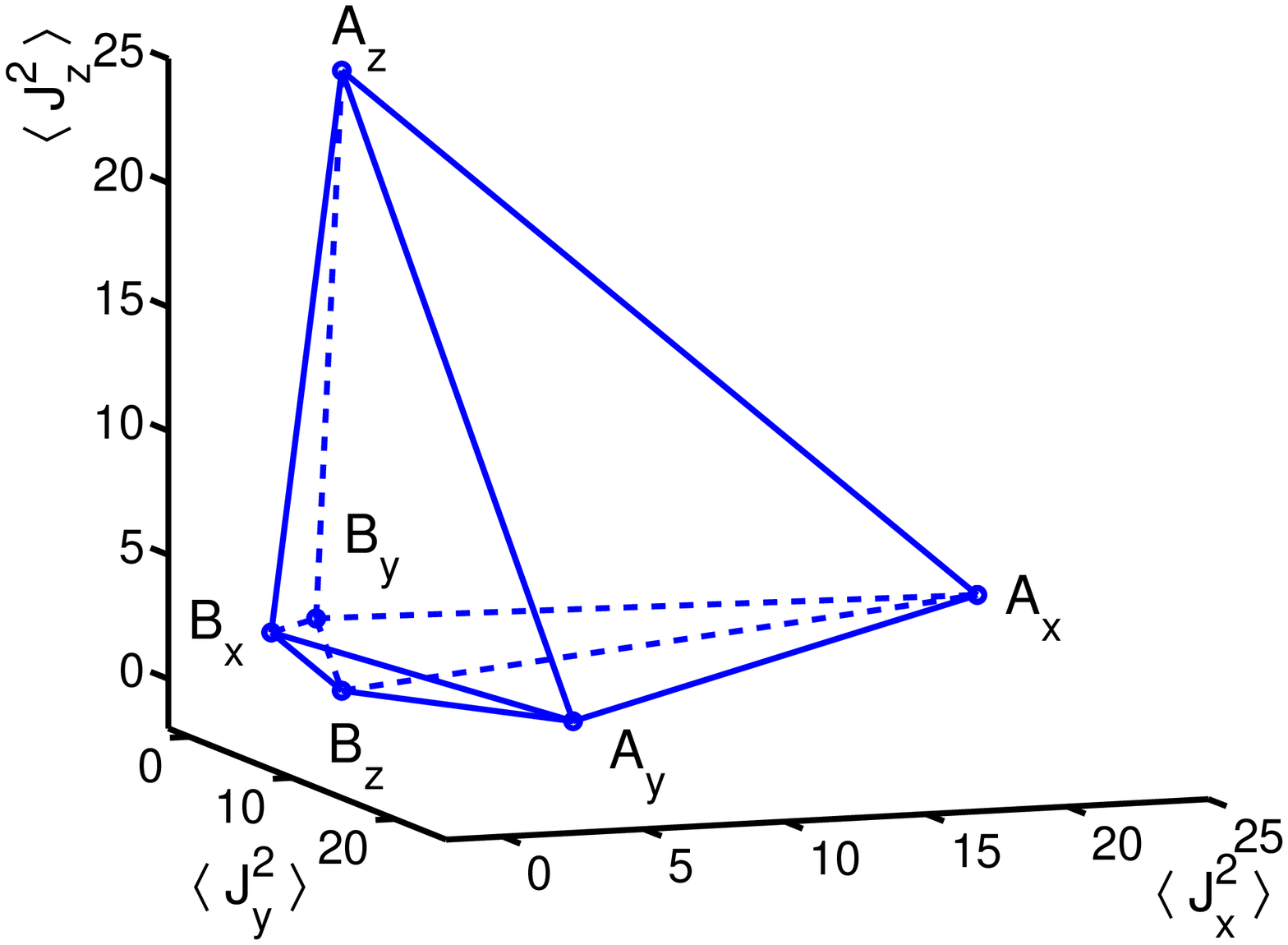}}
\centerline{ \epsfxsize3.3in \epsffile{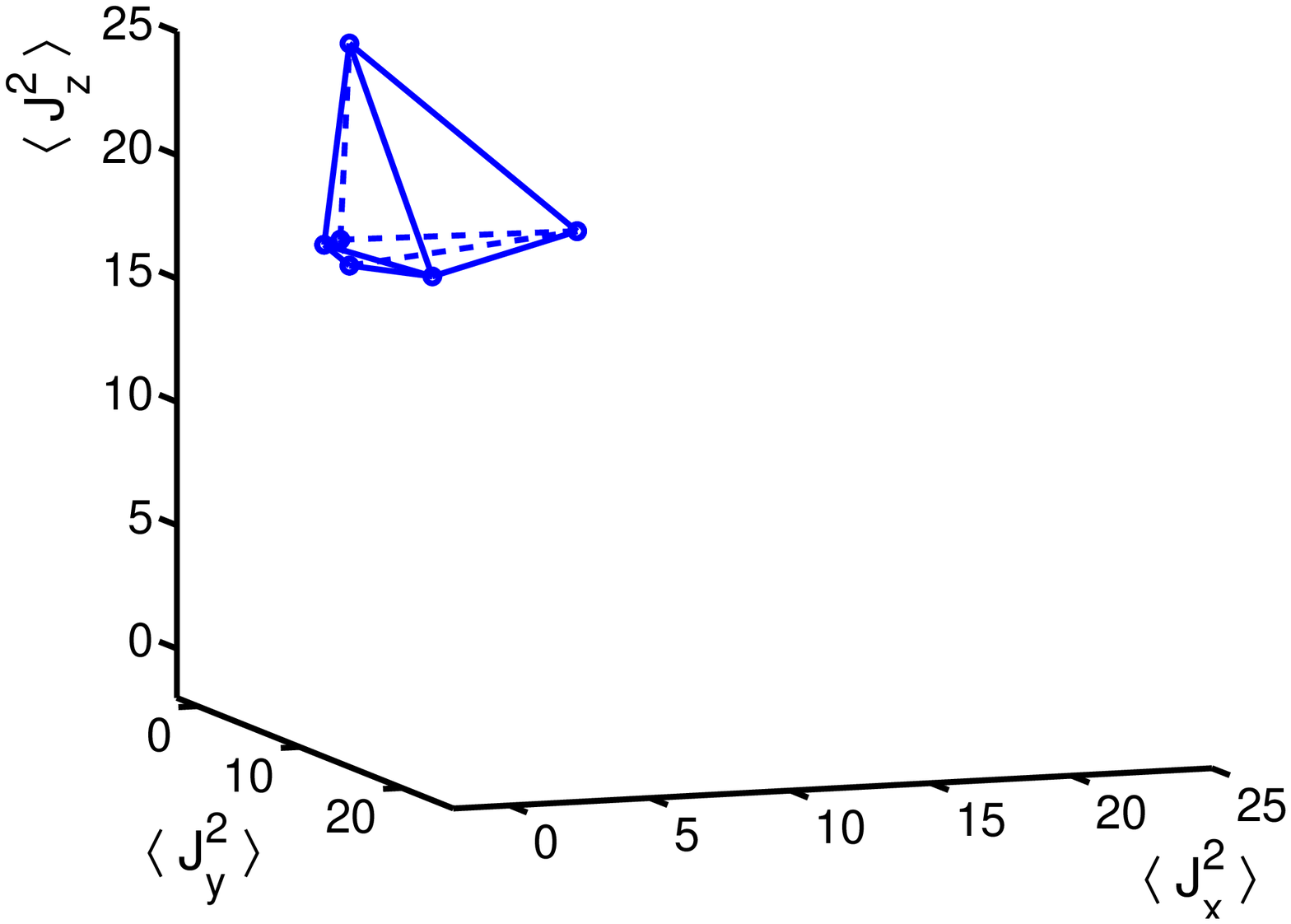}}
\caption{(a) The polytope of separable states corresponding to
Eqs.~(\ref{Jxyzineq}) for $N=10$ and for $\vec{J}=0.$ The origin of
the coordinate system corresponds to a many-body singlet state. (b)
The same polytope for $\vec{J}=(0,0,4).$ Note that this polytope is
a subset of the polytope in (a).} \label{J2xyz}
\end{figure}

In Ref.~\cite{TK07}, we have presented a set of spin squeezing
inequalities for the detection of entanglement. We showed that these
inequalities are complete, in the sense that they can detect all
entangled states that can be detected by the knowledge of
$\mean{J_l}$ and $\va{J_l}$ for three orthogonal directions
$l=x,y,z.$ This completeness means the following: A state that is
not detected by the inequalities, cannot be distinguished from a
separable state by knowing  $\mean{J_l}$ and $\va{J_l}$ only.

In this paper we present extensions of this approach in several
directions. In Sec. II, we first present a detailed derivation of
the optimal spin squeezing inequalities from Ref.~\cite{TK07}. Then,
we consider the case when only the variances $\va{J_l}$ (and not the
mean values $\mean{J_l}$) are known, or when only the mean values
$\mean{J_l^2}$ are known. We derive the optimal spin squeezing
inequalities also for this case. In Sec. III, we consider the
case when $\mean{J_l}$ and $\va{J_l}$ are known not only in three
orthogonal directions, but in arbitrary directions. In this case, we
can reformulate the spin squeezing inequalities as inequalities for
correlation and covariance matrices. In Sec. IV, we compare our
optimal spin squeezing inequalities to other known entanglement
criteria. In Sec. V, we discuss the issue of detecting
entanglement of the multi-qubit quantum state vs. detecting
entanglement in the reduced two-qubit density matrix. Finally, in
Sec. VI we apply our inequalities to the investigation of spin
models. We have shown already in Ref.~\cite{TK07} that the spin
squeezing inequalities can detect bound entanglement (a weak form of
entanglement, which is at the heart of many fundamental problems in
entanglement theory) in such models. Here, we present more examples
for the applicability of the spin squeezing inequalities.

\section{The optimal spin squeezing inequalities}

Our aim is to characterize the separable states in terms of the
values of  $\mean{J_l}$ and $\va{J_l}.$ Note that the knowledge of
$\mean{J_l}$ and $\va{J_l}$ is equivalent to the knowledge of
$\mean{J_l}$ and $\mean{J_l^2}.$ We now present our main result from
Ref.~\cite{TK07}:
\\
{\bf Observation 1.} {\it Let us assume that for a physical system
the values of
\be\vec{J}:=(\exs{J_x},\exs{J_y},\exs{J_z})\label{J}\ee and
\be\vec{K}:=(\exs{J_x^2},\exs{J_y^2},\exs{J_z^2})\label{K}\ee are
known. For separable states all the following inequalities are
fulfilled:}

\begin{subequations}
\begin{eqnarray}
\exs{J_x^2}+\exs{J_y^2}+\exs{J_z^2} &\le& \tfrac{N(N+2)}{4},
\label{theorem1a}
\\
\va{J_x}+\va{J_y}+\va{J_z} &\ge& \tfrac{N}{2},
\label{Jxyzineq_singlet}
\\
\exs{J_k^2}+\exs{J_l^2}-\tfrac{N}{2} &\le&  (N-1)\va{J_m},
\label{Jxyzineq_spsq2}
\\
(N-1)\left[\va{J_k}+\va{J_l}\right] &\ge&
\exs{J_m^2}+\tfrac{N(N-2)}{4}, \label{Jxyzineq_spsq3} \;\;\;\;\;\;
\end{eqnarray}
\label{Jxyzineq}
\end{subequations}
{\it where $k,l,m$ take all the possible permutations of $x,y,z.$
While Eq.~(\ref{theorem1a}) is valid for all quantum states,
violation of any of
Eqs.~(\ref{Jxyzineq_singlet}-\ref{Jxyzineq_spsq3}) implies
entanglement.}

{\it Proof. }  The variance, defined as
$\va{A}:=\exs{A^2}-\exs{A}^2,$ is concave in the state, that is, if
$\vr= p \vr_1 + (1-p) \vr_2$, then $\va{A}_\vr \geq
p\va{A}_{\vr_1}+(1-p)\va{A}_{\vr_2}.$ Thus, it suffices to prove
that the inequalities of Observation 1 are satisfied by pure product
states. Based on the theory of angular momentum, inequality
Eq.~(\ref{theorem1a}) is valid for all quantum states and the
equality holds for states of the symmetric subspace. However, for
separable states it can be proved easily without this knowledge
using that for such states \cite{GT05} \be
\exs{\sigma_x^{(i)}\sigma_x^{(j)}}+
\exs{\sigma_y^{(i)}\sigma_y^{(j)}}+
\exs{\sigma_z^{(i)}\sigma_z^{(j)}} \leq 1. \ee For
Eq.~(\ref{Jxyzineq_singlet}) one first needs that for product states
\be \va{J_k}= \tfrac{N}{4}-\tfrac{1}{4}\sum_i
\exs{\sigma_k^{(i)}}^2\ee holds. Then, for a product state one has
\be \va{J_x}+\va{J_y}+\va{J_z}=\tfrac{3N}{4}-\tfrac{1}{4}\sum_k
x_k^2+y_k^2+z_k^2. \label{proofJxyz} \ee Here
$x_i:=\exs{\sigma_x^{(i)}},$ $y_i:=\exs{\sigma_y^{(i)}},$ and
$z_i:=\exs{\sigma_z^{(i)}}.$ Knowing that $x_i^2+y_i^2+z_i^2\le 1,$
the right hand side of Eq.~(\ref{proofJxyz}) is bounded from below
by $\tfrac{N}{2}.$

Concerning Eq.~(\ref{Jxyzineq_spsq2}), we have to show that \be
\mathfrak{Y}:=(N-1)\va{J_x}+\tfrac{N}{2}-\exs{J_y^2}-\exs{J_z^2}\geq
0.\ee This can be written as \bea
\mathfrak{Y}&=&(N-1)[\tfrac{N}{4}-\tfrac{1}{4}\sum_i x_i^2] -
\tfrac{1}{4}\sum_{i \neq j} (y_i y_j + z_i z_j) \nonumber\\&=&
(N-1)[\tfrac{N}{4}-\tfrac{1}{4}\sum_i x_i^2]- \tfrac{1}{4}[(\sum_{i}
y_i)^2 + (\sum_{i} z_i)^2] \nonumber\\&+&
\tfrac{1}{4}\sum_i(y_i^2+z_i^2). \eea Using \be(\sum_i s_i)^2 \le N
\sum_i s_i^2,\label{bound} \ee and the normalization of the Bloch
vector, it follows that \be \mathfrak{Y}\geq \tfrac{N-1}{4}\sum_i
(1-x_i^2-y_i^2-z_i^2)\geq 0. \ee

Eq.~(\ref{Jxyzineq_spsq3}) can be proved in a similar way. We have
to show that
\be\mathfrak{Z}:=(N-1)\left[\va{J_k}+\va{J_l}\right]-\exs{J_m^2}-\tfrac{N(N-2)}{4}\geq
0.\ee This can be proved by rewriting $\mathfrak{Z}$ with the
individual spin coordinates and using Eq.~(\ref{bound}):
\begin{eqnarray}
\mathfrak{Z}&=&(N-1)[\tfrac{N}{4}-\tfrac{1}{4}\sum_i x_i^2 + y_i^2]
- \tfrac{1}{4}\sum_{i \neq j} z_i z_j
\nonumber\\
&\geq&\tfrac{N-1}{4}\sum_i(1-x_i^2-y_i^2-z_i^2)\geq 0.
\end{eqnarray}
\qed

For any value of $\vec{J}$ the eight inequalities
Eqs.~(\ref{Jxyzineq}) define a polytope in the three-dimensional
$(\exs{J_{x}^2},\exs{J_{y}^2},\exs{J_{z}^2})$-space. Observation 1
states that separable states lie inside this polytope. The polytope
is depicted in Figs.~\ref{J2xyz}(a,b) for different values for
$\vec{J}.$ It is completely characterized by its extremal points.
Direct calculation shows that the coordinates of the extreme points
in the $(\exs{J_{x}^2},\exs{J_{y}^2},\exs{J_{z}^2})$-space are
\begin{align}
A_x &:=\left[ \frac{N^2}{4}-\kappa(\exs{J_y}^2+\exs{J_z}^2),
\frac{N}{4}+\kappa\exs{J_y}^2, \frac{N}{4}+\kappa\exs{J_z}^2
\right], \nonumber
\\
B_x&:=\left[ \exs{J_x}^2+\frac{\exs{J_y}^2+\exs{J_z}^2}{N},
\frac{N}{4}+\kappa \exs{J_y}^2, \frac{N}{4}+\kappa\exs{J_z}^2
\right], \nonumber
\end{align}
where $\kappa:=(N-1)/N.$ The points $A_{y/z}$ and $B_{y/z}$ can be
obtained in an analogous way. Note that the coordinates of the
points $A_k$ and $B_k$ depend nonlinearly on $\exs{J_k}.$

\begin{figure}
\centerline{ \epsfxsize3.3in \epsffile{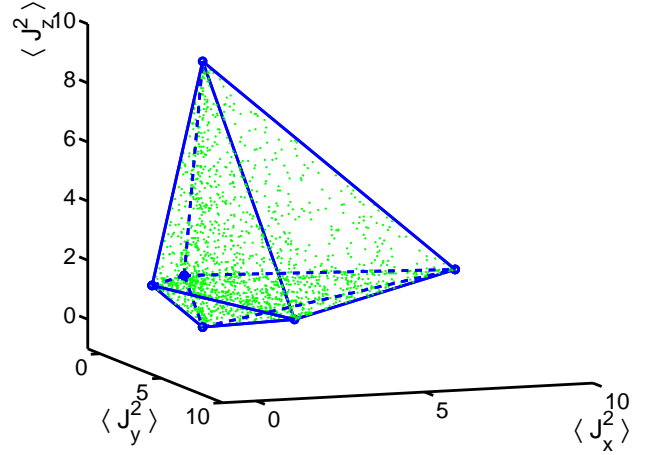}}
\caption{The polytope of separable states corresponding to
Eqs.~(\ref{Jxyzineq}) for $N=10$ and for $\vec{J}=0.$ The points
corresponding to random separable states fill the polytope. }
\label{random}
\end{figure}

One might ask whether all points inside the polytope correspond to
separable states. This would imply that the criteria of Observation
1 are complete, that is, if the inequalities are satisfied, then the
first and second moments of $J_k$ do not suffice to prove
entanglement. In other words, it is not possible to find criteria
detecting more entangled states based on these moments. Due to the
convexity of the set of separable states, it is enough to
investigate the extremal points:
\\
{\bf Observation 2.}
{\it (i) For any value of $\vec{J}$ there are
separable states corresponding to $A_k.$\\
(ii) If we define $J:=N/2,$ \be
c_x:=\sqrt{1-{(\exs{J_y}^2+\exs{J_z}^2)}/{J^2}}\ee and
$p:=[1+\exs{J_x}/(J c_x)]/2$ and if then $Np$ is an integer, then
there is also a separable state corresponding to $B_x.$ Similar
statements hold for $B_y$ and $B_z.$ Note that this condition is
always fulfilled, if $\vec{J}=0$ and $N$ is even.
\\
(iii) There are always separable states corresponding to points
$B_k'$ such that their distance from $B_k$ is smaller than
$\tfrac{1}{4}.$ In the limit $N\rightarrow \infty$ for a fixed
normalized angular momentum $\vec{j}:=\vec{J}/(N/2),$ the difference
between the volume of polytope of Eqs.~(\ref{Jxyzineq}) and the
volume of set of points corresponding to separable states decreases
with $N$ at least as $\Delta V/V \propto N^{-2},$ hence in the
macroscopic limit the characterization is complete. }

{\it Proof.} A separable state corresponding to $A_x$ is
\begin{equation}
{\vr}_{A_x}:=p(\ketbra{\psi_+})^{\otimes N}+
(1-p)(\ketbra{\psi_-})^{\otimes N}. \label{Ax}
\end{equation}
Here $\ket{\psi_{+/-}}$ are the single qubit states with Bloch
vector coordinates $(\exs{\sigma_x},\exs{\sigma_y},\exs{\sigma_z})=
(\pm c_x,\exs{J_y}/J,\exs{J_z}/J).$ If $M:=Np$ is an integer, we can
also define the state corresponding to the point $B_x$ as \be
\ket{\phi_{B_x}}:=\ket{\psi_+}^{\otimes M}\otimes
\ket{\psi_-}^{\otimes (N-M)}. \label{Bx} \ee Since there is a
separable state for each extreme point of the polytope, for any
internal point a corresponding separable state can be obtained by
mixing the states corresponding to the extreme points. It is
instructive to demonstrate this through a simple numerical
experiment. Fig.~\ref{random} shows that for $N=10$ and $\vec{J}=0$
random separable states indeed fill the polytope.

If $M$ is not an integer, we can approximate $B_x$ by taking $m:=
M-\varepsilon$ as the largest integer smaller than $M,$ defining
\begin{eqnarray}\vr'&:=&(1-\varepsilon) (\ketbra{\psi_+})^{\otimes m}
\otimes (\ketbra{\psi_-})^{\otimes(N-m)}  \nonumber \\&+&\varepsilon
(\ketbra{\psi_+})^{\otimes (m+1)} \otimes (\ketbra{\psi_-})^{\otimes
(N-m-1)}.\nonumber\\
\end{eqnarray}
This state has the same
coordinates as $B_x,$ except for the value of $\exs{J_x^2},$ where
the difference is $ c_x^2 (\varepsilon- \varepsilon^2) \le 1/4.$ The
dependence of $\Delta V/V$ on $N$ can be studied by considering the
polytopes in the $(\exs{J_{x}^2},\exs{J_{y}^2},\exs{J_{z}^2})$-space
corresponding to $\exs{J_k}=j_k\times \tfrac{N}{2},$ where $j_k$ are
the normalized angular momentum coordinates. As $N$ increases, the
distance of the points $A_k$ to $B_k$ scales as $N^2,$ hence the
volume of the polytope increases as $N^6.$ The difference between
the polytope and the points corresponding to separable states scales
like the surface of the polytope, hence as $N^4.$ \qed

Let us analyze now our optimal spin squeezing inequalities one by
one and define the corresponding facets of the polytope on
Fig.~\ref{J2xyz}(a). Eq.~(\ref{theorem1a}) corresponds to the facet
$A_x-A_y-A_z.$ As we discussed it is valid for all quantum states.
The symmetric states correspond to states on this facet and saturate
Eq.~(\ref{theorem1a}).

Eq.~(\ref{Jxyzineq_singlet}) has already been presented in
Ref.~\cite{GT04}. It corresponds to the facet $B_x-B_y-B_z.$ For even
$N,$ it is maximally violated by many-body singlets. For such states
\begin{eqnarray}
\vec{J}&:=&(0,0,0),\nonumber\\
\vec{K}&:=&(0,0,0). \label{Jsinglet}
\end{eqnarray}
That is, singlet states are states for which both the angular
momentum components and their variances are zero \cite{singlet}. For
large enough $N$ there are many states of this type. If we mix these
states, the mixture still maximally violates this inequality and
thus it is detected as entangled. This might be the reason that this
criterion can detect states that are very weakly entangled in the
sense that they are separable with respect to all bipartitions.

The violation of the criterion gives information about the number of
spins that are unentangled with the rest in the following sense
\cite{GT05}. Let us consider a pure state for which the first $M$
qubits are not entangled with other qubits while the rest of the
qubits are entangled with each other
\begin{equation}
\ket{\Psi}=(\otimes_{k=1}^{M} \ket{\psi_k}) \otimes \ket{\psi}_{M+1,...,N}.
\end{equation}
For such a state, based on the theory of entanglement detection with
uncertainties, we have \cite{unc}
\begin{equation}
\va{J_x}+\va{J_y}+\va{J_z} \ge \tfrac{M}{2}. \label{JxyzM}
\end{equation}
Let us consider now a mixed state $\vr:=\sum_k p_k \ketbra{\Psi_k}.$
If it violates Eq.~(\ref{JxyzM}) then at least one of the components
$\ketbra{\Psi_k}$ must have $M$ or more spins that are entangled
with other spins. If the left-hand side of Eq.~(\ref{JxyzM}) is
smaller than $\tfrac{1}{2}$ then the state cannot be created by
mixing states that have one or more unentangled spins.

Eq.~(\ref{Jxyzineq_spsq2}) corresponds to the facets $A_y-A_z-B_x,$
$A_x-A_z-B_y,$ and $A_x-A_y-B_z.$ All entangled symmetric Dicke
states violate this criterion \cite{EngDicke}. This can be seen as
follows. An $N$-qubit symmetric Dicke state with $m$ excitations is
defined as \cite{SG03}
\begin{equation}
\ket{m,N}:=\bigg(\begin{array}{c}N \\
m\end{array}\bigg)^{-\frac{1}{2}}\sum_k P_k
(\ket{1_1,1_2,...,1_m,0_{m+1},...,0_N}), \label{sd}
\end{equation}
where $\{P_k\}$ is the set of all distinct permutations of the
spins. $\ket{1,N}$ is the well known $N$-qubit W state. For states
of the form Eq.~(\ref{sd})
\begin{eqnarray}
\vec{J}&=&(0,0,m-\tfrac{N}{2}),\nonumber\\
\vec{K}&=&\big[\tfrac{N}{4}+\tfrac{m(N-m)}{2},
\tfrac{N}{4}+\tfrac{m(N-m)}{2},(m-\tfrac{N}{2})^2\big].
\label{jxyzdicke}
\end{eqnarray} Using Eqs.~(\ref{jxyzdicke}) one finds that
Eq.~(\ref{Jxyzineq_spsq2}) is violated by all Dicke states expect
for the non-entangled ones with $m=0$ and $m=N.$ For even $N,$ it is
maximally violated by the symmetric Dicke state
$\ket{\tfrac{N}{2},N}.$

Finally, Eq.~(\ref{Jxyzineq_spsq3}) corresponds to the facets
$A_y-B_z-B_x,$ $A_x-B_z-B_y,$ and $A_z-B_x-B_y.$ Note that these
inequalities detect the singlet state with Eq.~(\ref{Jsinglet}) as
entangled.

Now we can ask the question, what happens if we only know $\vec{K}$
from Eq.~(\ref{K}) and not $\vec{J}$ from Eq.~(\ref{J}). Can we
construct a polytope of the separable states similar to Observation
1? Similarly, we can consider the case that we know the variances
$[\va{J_{x}},\va{J_{y}},\va{J_{z}}],$ but not $\vec{J}.$ The
following observation gives the answer.

{\bf Observation 3.} {\it (i) Let us consider the set of points
corresponding to separable states for even $N$ in the
$(\ex{J_x^2},\ex{J_y^2},\ex{J_z^2})$-space \emph{without}
constraining the value of $\vec{J}.$ This set is the polytope from
Observation 1 for $\vec{J}=0,$ also shown in Fig.~\ref{J2xyz}(a).
\\
(ii) Also, the set of points corresponding to separable states in
the $[\va{J_{x}},\va{J_{y}},\va{J_{z}}]$-space is the same polytope.
That is, Fig.~\ref{J2xyz}(a) gives also the right polytope if the
labels of the axes are changed from $\exs{J_l^2}$ to $\va{J_l}.$ }

{\it Proof.} For the first part, it can be directly seen that
Eqs.~(\ref{Jxyzineq}) are least restrictive for $\vec{J}=0,$ for
other $\vec{J}$ the polytope is strictly smaller. For the second
part, note that based on Eqs.~(\ref{Jxyzineq}) the points
corresponding to separable states must be within the same polytope
shown in Fig.~\ref{J2xyz}(a), even if we change the labels from
$\exs{J_l^2}$ to $\va{J_l}.$ It is not clear, however, that the set
of separable states is convex in the
$[\va{J_{x}},\va{J_{y}},\va{J_{z}}]$-space. Thus, we have to show
that for each separable state $\vr$ with $\exs{J_l^2}=S_l$ for
$l=x,y,z,$ there is a separable state $\tilde{\vr}$ for which
$\va{J_l}=S_l.$ Let us use the decomposition $\vr = \sum p_k
\vr_{k}$ where $ \vr_{k}=\vr_k^{(1)}\otimes \vr_k^{(2)}\otimes ...
\otimes \vr_k^{(N)} $ are product states. Then, such a $\tilde{\vr}
:= \sum p_k \tilde{\vr}_{k}$ can be obtained by mixing
\begin{equation}
\tilde{\vr}_{k}:= \tfrac{1}{4} \big( \vr_{k} +
J_x\vr_{k}J_x+J_y\vr_{k}J_y+J_z\vr_{k}J_z \big).
\end{equation}
The state $\tilde{\vr}$ has the same $\exs{J_l^2}$ as $\vr.$
However, the value of $\exs{J_l}^2$ is zero, hence
$\va{J_l}_{\tilde{\vr}}=\exs{J_l^2}_\vr.$ \qed

\section{Optimal spin squeezing inequalities for the correlation matrix}

We discuss some further features of our spin squeezing inequalities.
One can ask what happens, if not only $\exs{J_k}$ and $\exs{J^2_k}$
for $k=x,y,z$ are known, but $\exs{J_{i}}$ and $\exs{J^2_{i}}$ in
arbitrary directions $i$. We will now first show how to find the optimal
directions $x',y',z'$ to evaluate Observation~1.

Knowledge of $\exs{J_{i}}$ and $\exs{J^2_{i}}$ in arbitrary
directions is equivalent to the knowledge of the vector $\vec{J},$
the correlation matrix $C$ and the covariance matrix $\gamma,$
defined as
\cite{ghge,DU06,Rivas}
\begin{eqnarray} C_{kl}&:=&\tfrac{1}{2}\exs{J_k
J_l+J_l J_k},\;\;\; \nonumber\\ \gamma_{kl}&:=&C_{kl} -
\exs{J_k}\exs{J_l},
\end{eqnarray}
for $k,l=x,y,z.$ When changing the coordinate system to $x',y',z',$
vector $\vec{J}$ and the matrices $C$ and $\gamma$ transform as
$\vec{J} \mapsto O \vec{J},$ $C\mapsto O C O^T$ and $\gamma \mapsto
O \gamma O^T$ where $O$ is an orthogonal $3\times3$-matrix. Looking
at the inequalities of Observation 1 one finds that the first two
inequalities are invariant under a change of the coordinate system.
Concerning Eq.~(\ref{Jxyzineq_spsq2}), we can reformulate it as \be
\exs{J^2_i}+\exs{J^2_j}+\exs{J^2_k}-\tfrac{N}{2} \leq (N-1) \va{J_k}
+ \exs{J^2_k}. \ee Then, the left hand side is again invariant under
rotations, and we find a violation of Eq.~(\ref{Jxyzineq_spsq2}) in
some direction if the minimal eigenvalue of \be
\mathfrak{X}:=(N-1)\gamma+C \ee is smaller than
$\trace(C)-\tfrac{N}{2}.$ Similarly, we find a violation of
Eq.~(\ref{Jxyzineq_spsq3}) if the largest eigenvalue of
$\mathfrak{X}$ exceeds $(N-1)\trace(\gamma)-N(N-2)/4.$ Thus, the
orthogonal transformation that diagonalizes $\mathfrak{X}$ delivers
the optimal measurement directions $x',y',z'$
\cite{qubit4matlab}.\\
{\bf Observation 4.} {\it We can rewrite our conditions
Eqs.~(\ref{Jxyzineq}) in a form that is independent from the choice of the
coordinate system as}
\begin{subequations}
\begin{eqnarray}
\trace(C) &\le& \tfrac{N(N+2)}{4}, \label{theorem1aB}
\\
\trace(\gamma) &\ge& \tfrac{N}{2}, \label{Jxyzineq_singletB}
\\
\lambda_{\rm min} (\mathfrak{X}) &\ge& \trace(C)-\tfrac{N}{2},
\label{Jxyzineq_spsq2B}
\\
\lambda_{\rm max} (\mathfrak{X}) &\le&
(N-1)\trace(\gamma)-\tfrac{N(N-2)}{4}, \label{Jxyzineq_spsq3B}
\;\;\;\;\;\;
\end{eqnarray}
\label{JxyzineqB}
\end{subequations}
{\it where $\lambda_{\rm min}(A)$ and $\lambda_{\rm max}(A)$ are the
smallest and largest eigenvalues of matrix $A,$ respectively. If
Eqs.~(\ref{Jxyzineq}) are violated by a quantum state for any choice
of coordinate axes $x,y,$ and $z$ then Eqs.~(\ref{JxyzineqB}) are
also violated.}

The preceding Observation shows how the optimal directions $x,y,z$
can be chosen by diagonalizing the matrix $\mathfrak{X}.$ However,
if one diagonalizes $\mathfrak{X}$ and does not find a violation of
Eqs.~(\ref{JxyzineqB}), this does not a priori imply that $C,
\gamma$ and $\vec{J}$ are compatible with a separable state. The
knowledge that for the diagonal $\mathfrak{X}$ the off-diagonal
entries vanish gives some additional information about the state,
which may in principle be used as a signature for entanglement. We
will prove now, however, that this is not the case and that
diagonalizing $\mathfrak{X}$ and applying Eqs.~(\ref{JxyzineqB}) is
the best one can do if $C, \gamma$ and $\vec{J}$ are known.

\begin{figure}
\centerline{ \epsfxsize3.3in \epsffile{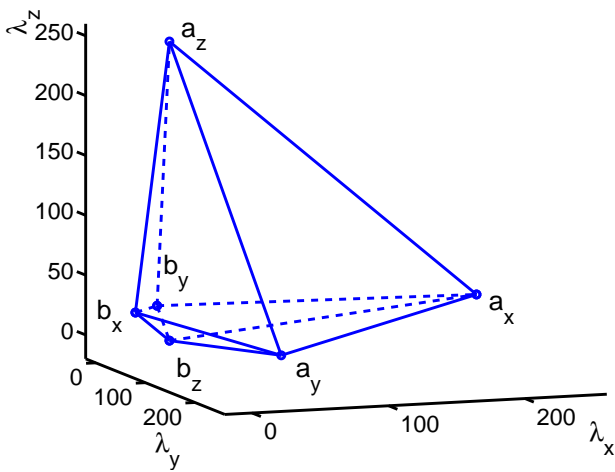}}
\caption{The polytope of separable states corresponding to
Eqs.~(\ref{JxyzineqC}) for the eigenvalues of $\mathfrak{X}$ for
$N=10$ and for $\vec{J}=0.$ Compare with Fig.~\ref{J2xyz}(a). }
\label{J2xyz_corrmat}
\end{figure}

Note first that Eqs.~(\ref{JxyzineqB}) contain the following variables:
the three eigenvalues of $\mathfrak{X},$ $\trace(C),$ and $\trace(\gamma).$
The latter two can be expressed with the trace of $\mathfrak{X},$ and
$\vec{J}$ as
\begin{eqnarray}
\trace(C)&=&\tfrac{1}{N}\trace(\mathfrak{X})+
\tfrac{N-1}{N}\vert\vec{J}\vert^2,\nonumber\\
\trace(\gamma)&=&\tfrac{1}{N}\trace(\mathfrak{X})-
\tfrac{1}{N}\vert\vec{J}\vert^2.
\end{eqnarray}
In this way, Eqs.~(\ref{JxyzineqB}) can be rewritten with the
eigenvalues of $\mathfrak{X}$ and $\vert\vec{J}\vert^2$ as
\begin{subequations}
\begin{eqnarray}
\trace(\mathfrak{X}) &\le&
\tfrac{N^2(N+2)}{4}-(N-1)\vert\vec{J}\vert^2, \label{theorem1aC}
\\
\trace(\mathfrak{X}) &\ge& \tfrac{N^2}{2}+\vert\vec{J}\vert^2,
\label{Jxyzineq_singletC}
\\
\lambda_{\rm min} (\mathfrak{X}) &\ge&
\tfrac{1}{N}\trace(\mathfrak{X})+
\tfrac{N-1}{N}\vert\vec{J}\vert^2-\tfrac{N}{2},
\label{Jxyzineq_spsq2C}
\\
\lambda_{\rm max} (\mathfrak{X}) &\le&
\tfrac{N-1}{N}\trace(\mathfrak{X})-
\tfrac{N-1}{N}\vert\vec{J}\vert^2-\tfrac{N(N-2)}{4},
\label{Jxyzineq_spsq3C} \;\;\;\;\;\;\;\;\;
\end{eqnarray}
\label{JxyzineqC}
\end{subequations}
For fixed $\vert\vec{J}\vert$ these equations describe a polytope in
the space of the three eigenvalues of $\mathfrak{X}$. The polytope
is shown in Fig.~\ref{J2xyz_corrmat}. The coordinates of the extreme
points in the $(\lambda_1,\lambda_2,\lambda_3)$ space of the
eigenvalues of $\mathfrak{X}$ are
\begin{eqnarray}
a_x:=\left[\tfrac{N^3}{4}-(N-1)\sum_k\exs{J_k}^2,\tfrac{N^2}{4},\tfrac{N^2}{4}
                 \right]\nonumber\\
\end{eqnarray}
and
\begin{eqnarray}
b_x:=\left[
                      \sum_k\exs{J_k}^2, \tfrac{N^2}{4}, \tfrac{N^2}{4}
                  \right].\nonumber\\
\end{eqnarray}
The other $a_k$ and $b_k$ points can be obtained by trivial
relabeling the coordinates.

Let us now show that in the large $N$ limit for any $\mathfrak{X}$
and $\vec J$ fulfilling Eqs.~(\ref{JxyzineqC}) there is a
corresponding quantum state. This would mean than the conditions
with $\mathfrak{X}$ and $\vec J$ are complete and there is not
another condition that could detect more entangled states based on
knowing $\mathfrak{X}$ and $\vec J$.

First, let us consider the case when $\vec{J}$ and $N$ fulfill the
conditions for completeness from Observation 2(ii), and there are
quantum states corresponding to $a_k$ and $b_k.$ The states
corresponding to $a_x$ and $b_x$ are $\vr_{A_x}$ and
$\vr_{B_x}:=\ketbra{\phi_{B_x}},$ respectively, defined in
Eqs.~(\ref{Ax}-\ref{Bx}). They are the same states that correspond to
the points $A_x$ and $B_x$ in Fig.~\ref{J2xyz}. The states
corresponding to the other extreme points can be obtained
straightforwardly from these formulas by relabeling the coordinates.
Note that all these states have a diagonal $\mathfrak{X}$ matrix.
Now, let us take a $\mathfrak{X}$ that fulfills
Eqs.~(\ref{JxyzineqC}) and diagonalize it, and denote it
$\mathfrak{X}_D$ after the diagonalization. Then, it is clear, that
$\mathfrak{X}_D$ can be obtained by "mixing" the $\mathfrak{X}$
matrices corresponding to $a_k$ and $a_k$ as \be \mathfrak{X}_D =
\sum_{l=a_x,a_y,a_z,b_x,b_y,b_z} p_l \mathfrak{X}_{l}, \ee where
$p_l>0$ and $\sum_k p_l=1.$ Note that "mixing" $\mathfrak{X}$
matrices is in general not equivalent to mixing the states, since
$\mathfrak{X}$ is a  nonlinear function of the state. However, for
all the states corresponding to $a_k$ and $b_k$ the vector $\vec{J}$
is the same and that all have diagonal $\mathfrak{X}$ matrices.
Therefore, the corresponding state is \be \vr_D =
\sum_{l=A_x,A_y,A_z,B_x,B_y,B_z} p_l \vr_{l}. \ee Then, if $\vr_D$
is the quantum state corresponding to $\vec{J}$ and $\mathfrak{X}_D,$
then the quantum state corresponding to $\vec{J}$ and $\mathfrak{X}$
can be obtained from $\vr_D$ with coordinate rotations. Finally, if
$\vec{J}$ and $N$ are such that no quantum state exists that
corresponds to some of the points, then an argument similar to the
one in Observation 2 can be applied showing that at least there is a
quantum state corresponding to a point close to all $b_k's$ and
because of that in the macroscopic limit the characterization is
complete even in this case. Thus, we can
state:\\
{\bf Observation 5.}
{\it The criteria from Eqs.~(\ref{JxyzineqB})
are complete in the sense that under the conditions
of Observation 2 (ii) or for large $N$ they detect all
entangled states that can be detected knowing
$\vec{J}$ and the correlation matrix $C.$}

\section{Comparison with other spin squeezing criteria}

\begin{figure}
\centerline{ \epsfxsize3.3in
\epsffile{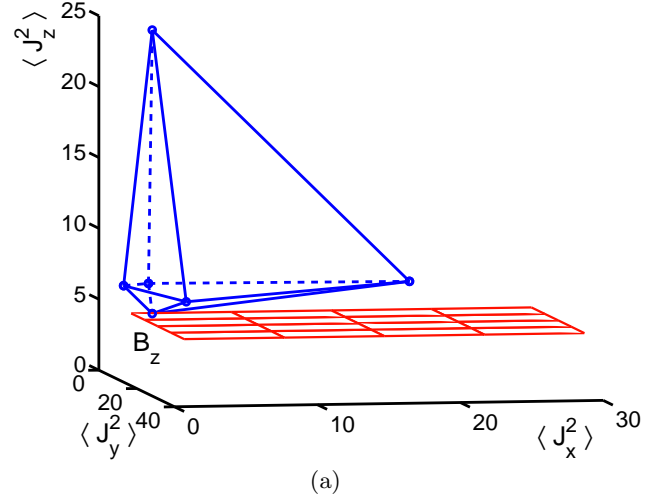}} \centerline{(a)}
\centerline{ \epsfxsize3.3in
\epsffile{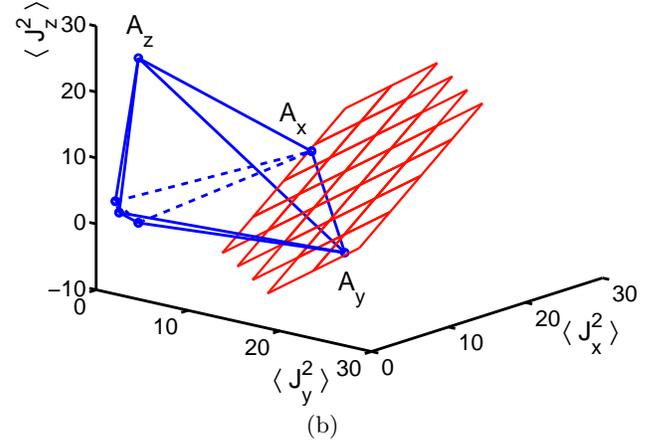}}
\centerline{(b)} \centerline{ \epsfxsize3.3in
\epsffile{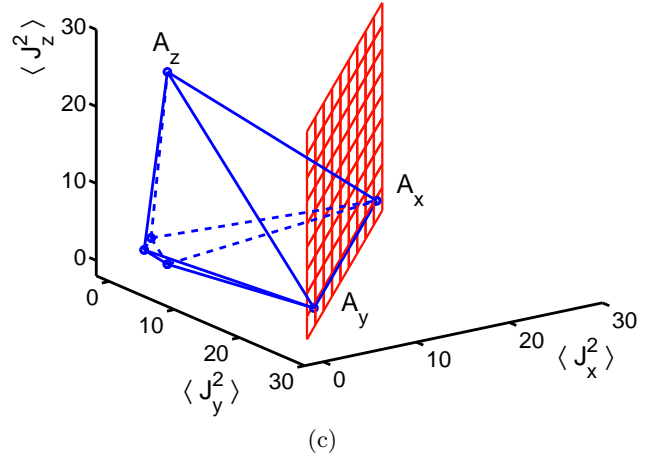}} \centerline{(c)}
\caption{(a) Comparison of the optimal spin squeezing inequalities
and original spin squeezing for $(k,l,m)=(x,y,z)$, $N=10$ and
$\vec{J}=(1,0,2).$ States detected by the latter are below the
horizontal plane. (b) Optimal spin squeezing inequalities and the
inequality Eq.~(\ref{TwoQubit}) for $N=10$ and $\vec{J}=(0,0,0).$
(c) Optimal spin squeezing and criterion Eq.~(\ref{Jx2Jy2}) for
$\vec{J}=(0,0,0).$} \label{J2xyz_comp}
\end{figure}

In this section we compare the optimal spin squeezing inequalities
Eqs.~(\ref{Jxyzineq}) to other spin squeezing criteria. First, let
us consider the original spin squeezing criterion
Eq.~(\ref{motherofallspinsqueezinginequalities}). This inequality is
satisfied by all points $A_k$ and $B_k,$ for $B_z$ even equality
holds. It is instructive to compare the region detected by
Eq.~(\ref{motherofallspinsqueezinginequalities}) to the region
detected by the optimal spin squeezing inequalities in the
$(\exs{J_x^2},\exs{J_y^2},\exs{J_z^2})$ space. For a fixed
$\vec{J},$ Eq.~(\ref{motherofallspinsqueezinginequalities})
corresponds to a horizontal plane in this space, shown in
Fig.~\ref{J2xyz_comp}(a).
Eq.~(\ref{motherofallspinsqueezinginequalities}) can be expressed in
a way that is independent from the choice of the coordinate system
\begin{equation}
\lambda_{\rm min}(\mathfrak{X})\ge \vert\vec{J}\vert^2
\label{spinsq_coord_indep}.
\end{equation}
Eq.~(\ref{spinsq_coord_indep}) is violated if
Eq.~(\ref{motherofallspinsqueezinginequalities}) is violated for an optimal
choice of coordinate axes $x,y,$ and $z.$

For a state of many particles that has almost a maximal spin in some
direction, the standard spin squeezing inequality
Eq.~(\ref{motherofallspinsqueezinginequalities}) is equivalent to
our optimal spin squeezing inequality Eq.~(\ref{Jxyzineq_spsq2}). To
see that, let us now rewrite Eq.~(\ref{Jxyzineq_spsq2}) as
\begin{equation}
\frac{\va{J_z}} {\exs{J_x^2}+\exs{J_y^2}}  \ge
\frac{1}{N-1}-\frac{N}{2(N-1)(\exs{J_x^2}+\exs{J_y^2})}.
\label{spinsq_alternative}
\end{equation}
This can be transformed into
\begin{eqnarray}
\frac{\va{J_z}} {\exs{J_x}^2+\exs{J_y}^2}  &\ge&
\bigg[\frac{1}{N-1}-\frac{N}{2(N-1)(\exs{J_x^2}+\exs{J_y^2})}\bigg]\nonumber\\
&\times&\frac{\exs{J_x^2}+\exs{J_y^2}}{\exs{J_x}^2+\exs{J_y}^2}.
\label{spinsq_alternative2}
\end{eqnarray}
Let us assume that $N$ is large and the state has a large spin
pointing to the $x$ direction, that is,
$\exs{J_x^2}\approx\tfrac{N^2}{4}$ and $\exs{J_y^2}\approx
\tfrac{N}{4}.$ In this case $\exs{J_k^2}\approx\exs{J_k}^2,$ and the
right hand side of Eq.~(\ref{spinsq_alternative2}) is very close to
$\tfrac{1}{N}.$ At this point one can recognize
Eq.~(\ref{motherofallspinsqueezinginequalities}).

Ref.~\cite{GT04} presented a generalized spin squeezing inequality
for entanglement detection that is identical to
Eq.~(\ref{Jxyzineq_singlet}) of the optimal spin squeezing
inequalities. This inequality has been connected to susceptibility
measurements in solid state systems \cite{TK07,njp}.

Refs.~\cite{KC05,spexp} presented another generalized spin squeezing
inequality, for detecting two-qubit entanglement. According to this
criterion, for states with a separable two-qubit density matrix
\begin{equation}
\big(\exs{J_k^2}+\exs{J_l^2}-\tfrac{N}{2}\big)^2+(N-1)^2 \exs{J_m}^2
\le\big[\exs{J_m^2}+\tfrac{N(N-2)}{4}\big]^2 \label{TwoQubit}
\end{equation}
holds. This inequality is satisfied by all points $A_k$ and $B_k,$
while, when we choose $(k,l,m)=(x,y,z),$ for $A_x$ and $A_y$ even
equality holds. Fig.~\ref{J2xyz_comp}(b) shows the polytope of the
optimal spin squeezing inequality together with the plane
corresponding to the Eq.~(\ref{TwoQubit}) criterion. Any state below
the plane is detected as two-qubit entangled by
Eq.~(\ref{TwoQubit}). Note that Eq.~(\ref{Jxyzineq_spsq2}) of the
optimal spin squeezing inequalities detects all states detected by
Eq.~(\ref{TwoQubit}). Note, however, that Eq.~(\ref{TwoQubit}) detects
only
states with two-qubit entanglement while Eq.~(\ref{Jxyzineq_spsq2})
detects entangled states that can have separable two-qubit density
matrices. Eq.~(\ref{TwoQubit}) can be expressed in a coordinate
system independent way as
\begin{eqnarray}
&&\lambda_{\rm max}
\{[\tfrac{1}{2}N^2+1-2\trace(C)]C-(N-1)^2\gamma\}
\nonumber\\&&\;\;\;\;\;\;\;\;\;\;\;\;\;\;\;\;\;\;\;\;\;\;\;\;\le
\big[\tfrac{N(N-2)}{4}\big]^2-[\trace(C)-\tfrac{N}{2}]^2.\label{TwoQubit_coordindep}
\end{eqnarray}

For states of the symmetric subspace, Eq.~(\ref{TwoQubit}) can be
simplified to \cite{KC05,spexp}
\begin{equation}
\frac{4\va{J_n}}{N} \ge 1-\frac{4\exs{J_n}^2}{N^2}.
 \label{TwoQubitSym}
\end{equation}
Violation of Eq.~(\ref{TwoQubitSym}) for some coordinate axis $z$ is
a necessary and sufficient condition for two-qubit entanglement for
symmetric states \cite{twoqubitremark}. It can also be expressed in a form that is
independent of the choice of coordinate axes \cite{DU06}
\begin{equation}
\lambda_{\rm min}(\gamma+\tfrac{1}{N}\vec{J}\vec{J}^T)\ge
\tfrac{N}{4}. \label{TwoQubitSymCorr}
\end{equation}
This can be rewritten with $\mathfrak{X}$ as
\begin{equation}
\lambda_{\rm min}(\mathfrak{X})\ge \tfrac{N^2}{4}.
\label{TwoQubitSymCorr2}
\end{equation}

Finally, Refs.~\cite{GT06,spexp2} present a generalized spin
squeezing inequality detecting entanglement close to symmetric Dicke
states with $\exs{J_z}=0.$ For separable states we have
\begin{equation}
\exs{J_x^2}+\exs{J_y^2}\le \tfrac{N(N+1)}{4}. \label{Jx2Jy2}
\end{equation}
The inequality is satisfied by all points $A_k$ and $B_k,$ for $A_x$
and $A_y$ even equality holds. Fig.~\ref{J2xyz_comp}(c) shows the
polytope of the optimal spin squeezing inequality together with the
plane corresponding to the Eq.~(\ref{Jx2Jy2}) criterion. Any state
corresponding to points on the right-hand side of the vertical plane
is detected by Eq.~(\ref{Jx2Jy2}) as entangled. Eq.~(\ref{Jx2Jy2})
can be rewritten in a coordinate system independent way as
\begin{equation}
\lambda_{\rm min}(C)\ge \trace(C)-\tfrac{N(N+1)}{4}.
\label{TwoQubitSymCorr3}
\end{equation}

\section{Two-qubit entanglement vs. multipartite entanglement}

Next, it is interesting to ask what kind of entanglement is detected
by our criteria knowing that they contain only two-body correlation
terms of the form $\exs{\sigma_k^{(i)} \sigma_k^{(j)}}$ and do not
depend on higher order correlations. In fact, all quantities in our
inequalities can be evaluated based on the knowledge of the average
two-qubit density matrix \be \vr_{\rm
av2}:=\tfrac{1}{N(N-1)}\sum_{i\ne j}\vr_{ij}, \label{av2} \ee where
$\vr_{ij}$ is the reduced density matrix of qubits $i$ and $j.$
Do our
criteria simply detect entanglement of the two-qubit reduced state
of the density matrix? It will turn out that our criteria can detect
entangled states with separable two-qubit density matrices.

Our entanglement detection scheme is related to the
$N$-representability problem \cite{Nrepresentability}, i.e., to the
problem of finding multipartite quantum states that have a given set
of states as reduced states \cite{extension}. When detecting
entanglement based on $\vr_{av2},$ we ask: Is there a separable
$N-$qubit state that has $\vr_{av2}$ as the average two-qubit
reduced state. If the answer is no then we know that the system is
in an entangled state. Clearly, if $\vr_{av2}$ is entangled then
there is not an $N$-qubit separable quantum state that has it as a
reduced state.

Interestingly, it turns our that it is also possible that
$\vr_{av2}$ is separable, however, there is not an $N$-qubit
separable state that has $\vr_{av2}$ as reduced state. In this case,
we can conclude that the system is an entangled state even if
$\vr_{av2}$ is separable. A similar phenomenon can be observed in
the theory of cluster states \cite{hein}: These are states that are
defined as eigenstates of quasi-local operators. The total state is
uniquely determined by these quasi-local properties of the reduced
states, and it can happen that the reduced states are separable,
while the total state is highly entangled \cite{graphbeispiel}.

Let us elaborate this point a little bit more. If $\vr_{av2}$ is
separable and it is in the symmetric subspace then it can always be
written in the form \cite{KC05}
\begin{equation}
\vr_{av2}=\sum_k p_k \vr_k \otimes \vr_k
\end{equation}
with pure $\vr_k$ matrices. In this case, there always can be found
an $N$-qubit separable state that has this state as its reduced
state
\begin{equation}
\vr_{N}=\sum_k p_k \vr_l^{\otimes N}.
\end{equation}
Hence, we can make the following statement.\\
{\bf Observation 6.} {\it If any
generalized-spin squeezing criterion (i.e., an entanglement
criterion containing only $\exs{J_k}$ and $\exs{J_k^2}$) is violated
by a symmetric state then the state is entangled and the reduced
two-qubit density matrix of the state is also entangled. Note that
this has already been known for the case of the original spin
squeezing criterion
Eq.~(\ref{motherofallspinsqueezinginequalities}) \cite{WS03}.}\\

On the other hand, if $\vr_{av2}$ is not symmetric then it is
possible that $\vr_{av2}$ is separable but there is not an $N-$qubit
symmetric separable state that has it as its reduced two-qubit
state. This is the reason that entanglement conditions based on
$\vec{J}$ and $\vec{K}$ can detect entangled states with a separable
two-qubit density matrix. Next, we will examine all our entanglement
criteria from this point of view.

First, we will rewrite Eqs.~(\ref{Jxyzineq}) as equations for the
average two-qubit density matrix $\vr_{av2}.$ All expectation values
in these equations are computed for $\vr_{av2}.$
\begin{subequations}
\begin{eqnarray}
\Sigma&\le&1, \label{theorem1a_av2}
\\
\Sigma&\ge&-\tfrac{1}{N-1}+\tfrac{N}{N-1}
\sum_{i=x,y,z}\exs{\sigma_i\otimes\mathbbm{1}}^2 ,
\;\;\;\;\;\;\;\;\;\;\label{xxyyzz}
\\
\Sigma&\le&1+ N \bigg(\exs{\sigma_m \otimes
\sigma_m}-\exs{\sigma_m\otimes\mathbbm{1}}^2\bigg),
\;\;\;\;\;\;\;\;\;\;\label{xxyyzzz}
\\
\Sigma&\ge&-\tfrac{1}{N-1}+ \tfrac{N}{N-1} \bigg(\exs{\sigma_k
\otimes \mathbbm{1}}^2+\exs{\sigma_l \otimes
\mathbbm{1}}^2\bigg)\nonumber\\&&\;\;\;\;\;\;\;\;\;\;\;\;\;\;\;+\tfrac{N}{N-1}
\exs{\sigma_m \otimes
\sigma_m},\label{Jxyzineq_spsq3_av2}\;\;\;\;\;\;
\end{eqnarray}
\label{Jxyzineq_rhoav}
\end{subequations}
where \be \Sigma:=\exs{\sigma_x \otimes \sigma_x + \sigma_y \otimes
\sigma_y + \sigma_z \otimes \sigma_z}.\ee

The first of the four optimal spin squeezing inequalities,
Eq.~(\ref{theorem1a}), corresponds to Eq.~(\ref{theorem1a_av2}).
Again, this inequality is valid for any quantum state and cannot be
violated.

The second inequality, Eq.~(\ref{Jxyzineq_singlet}) corresponds to
Eq.~(\ref{xxyyzz}). It is known that Eq.~(\ref{Jxyzineq_singlet})
can detect states that have separable two-qubit density matrices as
have been shown in Ref.~\cite{GT05}. Such a state is, for example,
one of the many-body singlet states, \be \vr_{s}\propto
\lim_{T\rightarrow 0}\exp\big(-\frac{J_x^2+J_y^2+J_z^2}{T}\big). \ee
For this state $\exs{J_m^n}=0$ for any $m,n\ge 0.$ For such a state
for increasing $N$ the average two-qubit density matrix $\vr_{av2}$
becomes arbitrarily close to the totally mixed state. Thus,
Eq.~(\ref{xxyyzz}) is not a condition for detecting the entanglement
of $\vr_{av2}.$ Moreover, note that Eq.~(\ref{Jxyzineq_singlet}) can
even detect states that are separable with respect to all
bipartitions \cite{TK07}.

The third inequality Eq.~(\ref{Jxyzineq_spsq2}) corresponds to
Eq.~(\ref{xxyyzzz}). Let us consider the state \be \vr\propto
\exp\big(-\frac{7J_z^2-J_x^2-J_y^2}{T}\big) \ee for $N=8$ and $T=3$.
Direct calculation shows that this state is detected by
Eq.~(\ref{Jxyzineq_spsq2}) for $(k,l,m)=(x,y,z).$ Thus, again, this
is not a condition for the separability of the two-qubit density
matrix.

The fourth condition is Eq.~(\ref{Jxyzineq_spsq3}) which corresponds to
Eq.~(\ref{Jxyzineq_spsq3_av2}). It detects the singlet state
$\vr_{s}.$ This state has a separable two-qubit density matrix thus
Eq.~(\ref{Jxyzineq_spsq3}) is not a condition on the separability of
the reduced density matrix.

Let us consider now the original spin squeezing inequality
Eq.~(\ref{motherofallspinsqueezinginequalities}). It is known that
 the violation of this inequality
 implies two-qubit entanglement for symmetric states \cite{WS03}.
However, if the quantum state is not symmetric,
Eq.~(\ref{motherofallspinsqueezinginequalities}) can detect states
with separable two-qubit density matrices. For example, the
following state violates
Eq.~(\ref{motherofallspinsqueezinginequalities}), while it does not
have two-qubit entanglement: \be\vr_{sq} \propto
\exp\big(-\frac{2J_x^2-J_z}{T}\big) \ee for $N=8$ and $T=0.3$.

Finally, let us consider the generalized spin squeezing inequality
Eq.~(\ref{Jx2Jy2}). It can be proved that any state violating it has
two-qubit entanglement. This is because it can be rewritten with
expectation values computed for $\vr_{av2}$ as
\begin{equation}
\exs{\sigma_x\otimes\sigma_x}+\exs{\sigma_y\otimes\sigma_y}\le 1.
\label{Jx2Jy2_twoqubit}
\end{equation}
Any two-qubit state violating this criterion is entangled
\cite{GT05}.

\section{Spin systems giving violations for the optimal spin squeezing inequalities}

In the recent years, considerable effort has been made to create
large scale entanglement in various physical systems: In
Bose-Einstein condensates of two-state bosonic atoms \cite{SD01}, in
optical lattices of cold two-state atoms realizing the dynamics of
an Ising spin chain \cite{optlatt_gates,isingdyn_optlatt,WS01} and
in atomic clouds through the interaction with light and appropriately
chosen measurements \cite{KB98,HM04,HP06}. In the future, it is expected
that experimenters will also engineer the various ground states of
well known spin chains. Entanglement detection in such systems were
considered, for example, in Refs.~\cite{B03,GT04,AJ04,VC07}. Note
that there are methods available for measuring the variances of the
collective spin components of atomic systems through interaction
with light \cite{ER08,VC08}.

In the light of the experiments, we ask the question: Under what
circumstances are our optimal spin squeezing inequalities useful for
detecting entanglement in the sense that they outperform other spin
squeezing entanglement criteria? In this section, we will show that
our entanglement criteria are especially useful in situations in
which the state has a small or zero mean spin $\vec{J}$ and its
reduced average two-qubit density matrix $\vr_{\rm av2}$ is
separable.

\subsection{Ground state of spin systems}

These will be, on the one hand, one-dimensional spin chains. On the
other hand, we will consider spin systems corresponding to the
completely connected graph. We will consider the following
Hamiltonians:

First let us consider the Heisenberg chain with the Hamiltonian \be
H_{\rm H}:=\sum_k
\sigma_x^{(k)}\sigma_x^{(k+1)}+\sigma_y^{(k)}\sigma_y^{(k+1)}+\sigma_z^{(k)}\sigma_z^{(k+1)}.
\label{Hheisenberg} \ee Its ground state is a many-body singlet
state. Thus, the optimal spin squeezing inequality
Eq.~(\ref{Jxyzineq_singlet}) is ideal for its detection. Concerning
how other criteria can detect its ground state as entangled,
we can state the following:\\
{\bf Observation 7.} {\it The $T=0$ ground state of a spin system
with a Hamiltonian without an external field cannot be detected by
the original spin squeezing criterion
Eq.~(\ref{motherofallspinsqueezinginequalities}). The ground state
of a spin chain Hamiltonian without an external field cannot be
detected by the
Korbicz-Cirac-Lewenstein criterion Eq.~(\ref{TwoQubit}).}\\
{\it Proof. } The first statement is true since criterion
Eq.~(\ref{motherofallspinsqueezinginequalities}) cannot be used for
states with $\vec{J}=0$ since in
Eq.~(\ref{motherofallspinsqueezinginequalities}) one has to divide
with the length of the collective spin components. The other claim
can be proved noting that for large $N$ the two-qubit density matrix
$\vr_{\rm av2}$ of the ground state of spin chains without an
external field is unentangled. This can be seen realizing that for
the ground state of an $N$-qubit translationally invariant chain \be
\vr_{\rm
av2}=\tfrac{1}{N-1}\left(\vr_{12}+\vr_{13}+\vr_{14}+...+\vr_{1N}
\right),\ee where $\vr_{kl}$ is the reduced two-qubit matrix of
spins $k$ and $l.$ However, for a spin chain distant sites are less
and less correlated thus for large enough $k$ we have
$\vr_{1k}\approx\tfrac{1}{4}\openone.$ Hence, for large enough $N$
the reduced two-qubit matrix $\vr_{\rm av2}$ is very close to the
totally mixed state and it is separable \cite{ZH98}; thus, the state
is not detected by the Korbicz-Cirac-Lewenstein criterion
Eq.~(\ref{TwoQubit}). \qed

The Hamiltonian of the isotropic XY chain is \be H_{\rm XY}:=\sum_k
\sigma_x^{(k)}\sigma_x^{(k+1)}+\sigma_y^{(k)}\sigma_y^{(k+1)}.\label{HXY}\ee
This system is similar to Eq.(\ref{Hheisenberg}) from the point of
view of detecting its ground state by various entanglement criteria.
That is, $\vr_{\rm av2}$ is unentangled for this system and the spin
squeezing criterion Eq.~(\ref{motherofallspinsqueezinginequalities})
cannot detect its ground state. Moreover, its ground state is
detected by the optimal spin squeezing inequalities. While the XY
chain is exactly solvable \cite{XY}, the latter statement can be
understood based on simpler arguments using only qualitative
properties of the ground state. Let us consider a chain with a
periodic boundary condition. For the non-degenerate ground state of
the XY chain for even $N$ one has $\exs{J_z^2}=0$ since $J_z$
commutes with $H_{XY}.$ The nearest neighbor correlation is the
strongest, that is for the ground state \be
\exs{\sigma_l^{(m)}\sigma_l^{(n)}}=(-1)^{m-n}c_{l,D(m,n)}\ee for
$l=x,y$ where $D(m,n)$ is the distance of qubit $m$ and $n$, and
$c_{l,m}>0$ is a monotonous decreasing function of $m.$ Hence, due
to translational invariance, it follows that \be
\exs{J_l^2}=\frac{N}{4}+\frac{1}{4}\sum_{m \ne n}
\exs{\sigma_l^{(m)}\sigma_l^{(n)}}<
\frac{N}{4}-\frac{N}{2}\Delta_N\ee for $l=x,y,$ where
$\Delta_N:=\vert\exs{\sigma_l^{(1)}\sigma_l^{(2)}}\vert-\vert\exs{\sigma_l^{(1)}\sigma_l^{(3)}}\vert.$
Note that $\Delta_N$ converges to a non-zero value for
$N\rightarrow\infty.$
 Using these arguments, one can see that for any even $N$ the ground
state of the XY chain violates Eq.~(\ref{Jxyzineq_singlet}) and this
violation is of order $N$ in the large $N$ limit, that is, the
relative violation does not approach zero with increasing $N.$ Hence
it also follows that chains with odd $N$ must also violate
Eq.~(\ref{Jxyzineq_singlet}) in this limit.

The Hamiltonian \be H_{\rm
S}:=J_x^2+J_y^2+J_z^2=\frac{N}{4}+\frac{1}{2}\sum_{l=x,y,z}\sum_{
m>n}\sigma_l^{(m)}\sigma_l^{(n)} \label{Hsinglet} \ee corresponds to
a system that has a Heisenberg interaction between all spin pairs
and has a very degenerate ground state. The two-qubit density matrix
of its $T=0$ thermal ground state converges to the completely mixed
state as $N$ increases, thus for large enough $N$ it is separable
\cite{GT05}. With respect to other qualitative statements about
entanglement detection, it is similar to the Heisenberg chain.

The Hamiltonian of the Lipkin-Meshkov-Glick model is \cite{LMG}
\be H_{\rm LMG}:=-\frac{\lambda}{N}(J_x^2+\gamma
Jy^2)-hJ_z.\label{HLMG}\ee For $\lambda\ge 0,$ $\gamma=1$ and $h=0$
the ground state is an $N$-qubit symmetric Dicke states with
$\tfrac{N}{2}$ excitations. For $h\ne 0$ all the symmetric Dicke
states given in Eq.~(\ref{sd}) can be obtained as ground states of
the system. These, except for the trivial $\ket{0,N}=\ket{0000...}$
and $\ket{N,N}=\ket{1111...}$ states, all have entangled reduced
two-qubit density matrix. Using Eq.~(\ref{jxyzdicke}), one can show
that they are detected both by our optimal spin squeezing
inequalities and the Korbicz-Cirac-Lewenstein criterion
Eq.~(\ref{TwoQubit}). However, they are not detected by the original
spin-squeezing inequality as can be seen by substituting
Eqs.~(\ref{jxyzdicke}) into the original spin-squeezing inequality.
For $\lambda\le 0,$ $\gamma=1$ and $h=0$ the ground state is the
same as for the Hamiltonian Eq.~(\ref{Hsinglet}).

Finally, the summary of the results in this section is shown in Table
\ref{Htab}.

\begin{table}
\centerline{
\begin{tabular}{|c||c|c|c| }
\hline
  Hamiltonian &
  \begin{tabular}{c} Spin      \\ squeezing   \\ Eq.~(\ref{motherofallspinsqueezinginequalities}) \end{tabular} &
  \begin{tabular}{c} Korbicz-Cirac- \\ Lewenstein \\ Eq.~(\ref{TwoQubit})\end{tabular} &
  \begin{tabular}{c} Optimal spin \\ squeezing \\ Eqs.~(\ref{Jxyzineq})\end{tabular} \\
  \hline\hline
  \begin{tabular}{c} Heisenberg  \\  \text{chain}\end{tabular} & $-$ & $-$ & $+$  \\ \hline
  \begin{tabular}{c} XY \\  \text{chain}\end{tabular} & $-$ & $-$ & $+$  \\ \hline
  \begin{tabular}{c} Heisenberg model\\  \text{ fully connected}\end{tabular} & $-$ & $-$ & $+$  \\ \hline
  \begin{tabular}{c} $H_{\rm LMG}$ \\ $\lambda > 0$  \end{tabular} & $-$ & $+$ & $+$  \\ \hline
  \begin{tabular}{c} $H_{\rm LMG}$ \\ $\lambda < 0, h = 0 $ \end{tabular} & $-$ & $-$ & $+$  \\ \hline
\end{tabular}
} \caption{Table showing for several spin Hamiltonians which
entanglement condition can detect their $T=0$ thermal ground state
in the large particle number limit. For the Hamiltonians see text.
For the Lipkin-Meshkov-Glick model $\gamma=1$ is assumed.}
\label{Htab}
\end{table}

\subsection{Bound entanglement in spin chains}

\begin{figure}
{ \epsfxsize1.65in \epsffile{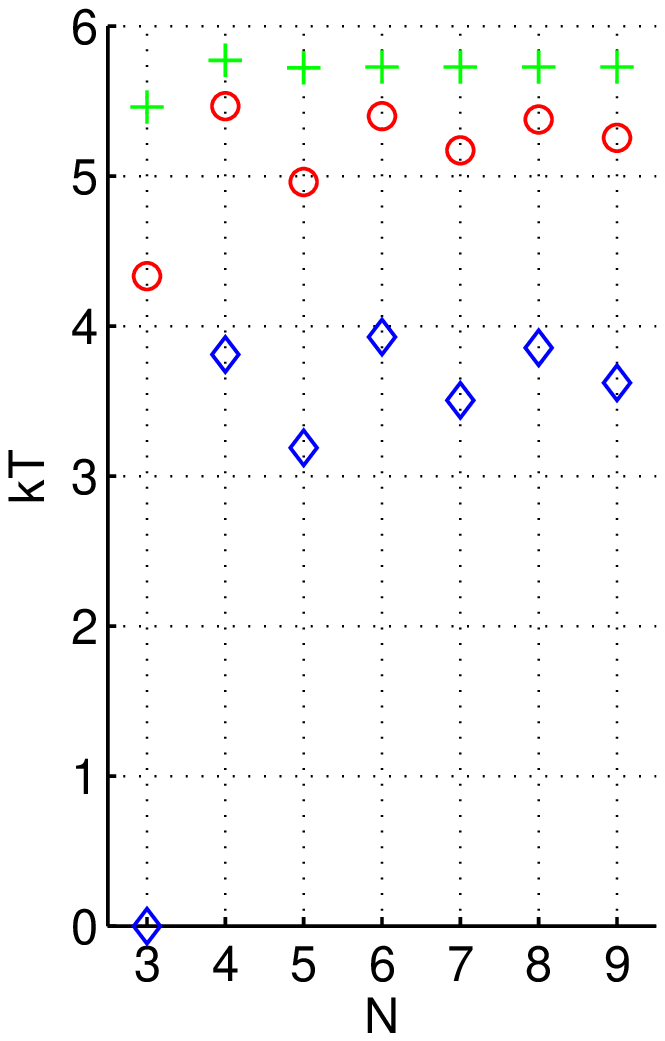}} {
\epsfxsize1.65in \epsffile{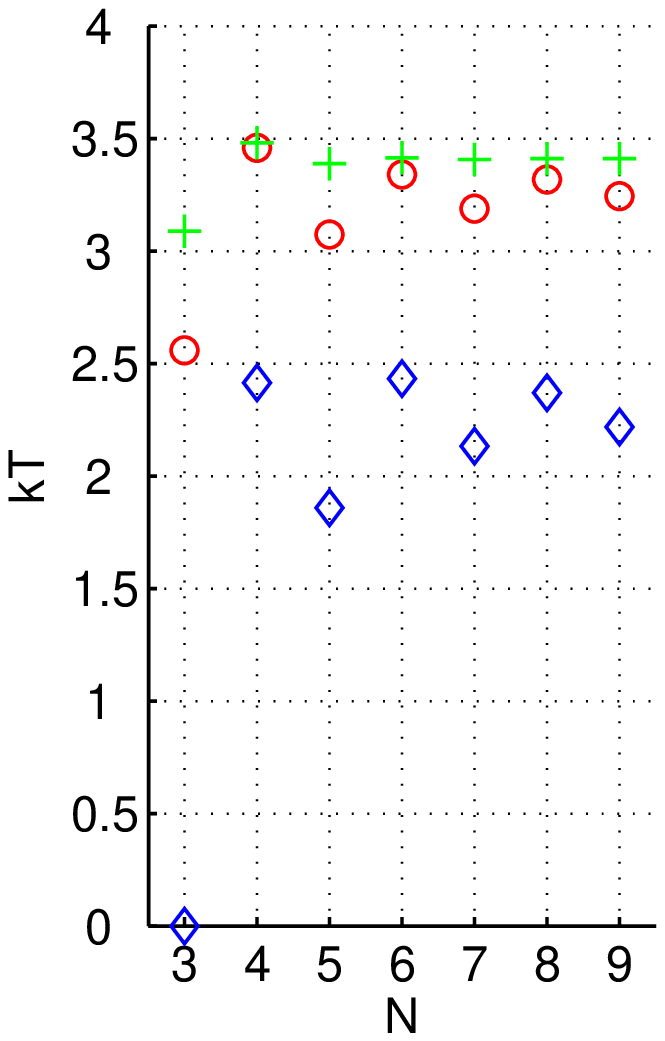}}
\caption{Comparison of the critical temperatures of separability
criteria for different site numbers in the Heisenberg- (left) and
the XY-model (right). ${T_c}$ of the spin squeezing inequality
Eq.~(\ref{Jxyzineq_singlet}) ($+$) is higher than the critical
temperatures of the PPT ($\bigcirc$) criterion \cite{ppt} or the CCNR
(${\diamond}$) criterion \cite{ccnr}.} \label{spinchain_3to9}
\end{figure}

Next, we study spin models in thermal equilibrium. We give the
threshold temperatures for various spin models for the PPT criterion
\cite{ppt} and for our optimal spin squeezing inequalities
Eqs.~(\ref{Jxyzineq}). These temperatures are defined as the values,
below which the spin squeezing inequalities are violated of the
state becomes NPT with respect to at least one partition. The
results are given in Table~\ref{tb}. The systems considered are the
Heisenberg chain and the XY chain defined in
Eqs.~(\ref{Hheisenberg},\ref{HXY}), the Heisenberg system on a fully
connected graph with the Hamiltonian Eqs.~(\ref{Hsinglet}), the XY
system on a fully connected graph with the Hamiltonian $H_{\rm LMG}$
defined in Eq.~(\ref{HLMG}) for $h=0, \gamma=1$ and $\lambda=-1,$
and the antiferromagnetic Ising spin chain in a transverse field
defined as \be H_{\rm I}:=\sum_k
\sigma_z^{(k)}\sigma_z^{(k+1)}+B\sum_k\sigma_x^{(k)}.\label{Hising}\ee
The thermal state of the system is computed as $\rho_{\rm
th}\propto\exp(-\tfrac{H}{kT})$ with $k=1.$ In many cases, the
temperature bound for the PPT criterion is lower than for our spin
squeezing criterion. This means that there is a temperature range in
which the quantum state has a positive partial transpose with
respect to all bipartitions while it is still detected as entangled.
Such quantum states are bound entangled and since all bipartitions
are PPT, no entanglement can be distilled from them with local
operations and classical communications even if arbitrary number of
parties are allowed to join \cite{NPTbound}. In particular, the
results show that Eqs.~(\ref{Jxyzineq}) can detect fully PPT bound
entanglement in Heisenberg and XY chains, moreover, in Heisenberg
and XY systems on a completely connected graph. Note that the bound
temperature for the optimal spin squeezing inequalities for the
Heisenberg model on a fully connected graph is $T_c\approx N$ for
large $N$ \cite{GT05}. On the other hand, our criteria do not seem
to detect fully PPT bound entanglement in Ising spin chains.
Finally, Fig.~\ref{spinchain_3to9} shows the results for the
Heisenberg and XY and chains, together with the bounds for the
computable cross norm or realignment (CCNR) criterion \cite{ccnr}.
The latter is often a good indicator of bound entanglement, however, in
these systems it does not detect bound entanglement.

\begin{table}
\centerline{ \begin{tabular}{|l l||c|c|c|c|c|c|c|}
  \hline
  &N & 3 & 4 & 5 & 6 & 7 & 8 & 9 \\
  \hline
  Heisenberg &\vline \;\;Eq.~(\ref{Jxyzineq_singlet}) & 5.46 & 5.77 & 5.72 & 5.73 & 5.73 & 5.73 & 5.73 \\
  chain &\vline \;\;PPT & 4.33 & 5.47 & 4.96 & 5.40 & 5.17 & 5.38 & 5.25 \\
  \hline
 XY  &\vline\;\;Eq.~(\ref{Jxyzineq_singlet}) & 3.09 & 3.48 & 3.39 & 3.41 & 3.41 & 3.41 & 3.41 \\
 chain &\vline\;\;PPT & 2.56 & 3.46 & 3.07 & 3.34 & 3.19 & 3.32 & 3.24
   \\
   \hline
    Heisenberg  &\vline\;\;Eq.~(\ref{Jxyzineq_singlet}) & 2.73 & 3.73 & 4.73 & 5.72 & 6.72 & 7.72 & 8.72 \\
 model f.c. &\vline\;\;PPT & 2.16 & 2.73 & 3.17 & 3.71 & 4.17 & 4.70 & 5.17
   \\
   \hline
    XY   &\vline\;\;Eq.~(\ref{Jxyzineq_singlet}) & 1.54 & 2.08 & 2.59 & 3.10 & 3.60 & 4.11 & 4.61 \\
 model f.c. &\vline\;\;PPT & 1.28 & 1.82 & 2.23 & 2.74 & 3.20 & 3.71 & 4.19
  \\
  \hline
   Ising chain   &\vline\;\;Eq.~(\ref{Jxyzineq_spsq2}) & 0.67 & 0.89 & 0.55 & 0.78 & 0.50 & 0.71 & 0.46 \\
 B=0.5 &\vline\;\;PPT & 1.08 & 1.26 & 1.17 & 1.26 & 1.21 & 1.26 & 1.22
  \\
  \hline
   Ising chain   &\vline\;\;Eq.~(\ref{Jxyzineq_spsq2}) & 1.22 & 1.29 & 1.14 & 1.17 & 1.10 & 1.11 & 1.08 \\
 B=1 &\vline\;\;PPT & 1.49 & 1.71 & 1.61 & 1.71 & 1.65 & 1.71 & 1.67
  \\
  \hline
   Ising chain   &\vline\;\;Eq.~(\ref{Jxyzineq_spsq2}) & 2.01 & 1.97 & 1.90 & 1.87 & 1.85 & 1.83 & 1.82 \\
 B=2 &\vline\;\;PPT & 2.15 & 2.43 & 2.30 & 2.43 & 2.36 & 2.43 & 2.38
   \\\hline
\end{tabular}} \caption{Critical temperatures for the PPT criterion  and Eqs.~(\ref{Jxyzineq})
for Heisenberg, XY and Ising spin chains of various size, and for
the Heisenberg any XY systems on a fully connected graph. For the
definitions of the Hamiltonians see text.} \label{tb}
\end{table}

\subsection{Bound entanglement in a nanotubular system}

\begin{figure}
\centerline{ \epsfxsize0.7\columnwidth
\epsffile{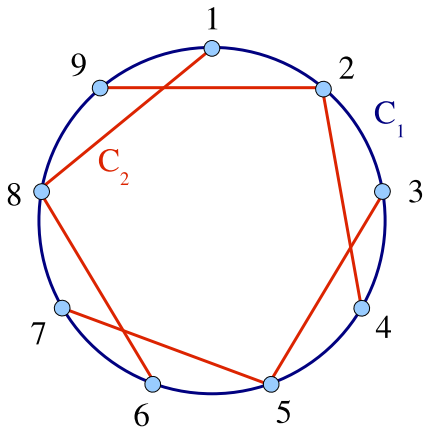}} \caption{ Schematic picture of
the ${\rm Na_2V_3O_7}$ system with coupling parameters ${C_1}$,
${C_2}$ of the nine spin-$\tfrac{1}{2}$ Heisenberg ring model.}
\label{tube_coupl}
\end{figure}

Let us finally investigate a finite system showing bound
entanglement at high temperatures. The nanotubular system ${\rm
Na_2V_3O_7}$ is a prominent example of a low-dimensional quantum
magnet. The compound was synthesized in 1999 by P. Millet et
al. \cite{tube_synth}, who also provided a detailed description of
its structure: Every nine ${\rm V^{4+}O_5}$ pyramids form a ring, by
sharing edges and corners, furthermore those rings accumulate to
nanotubes with Na atoms located in the center of and between them.
Due to the complex structure of this system some years passed, until
an effective model for the exchange interactions could be found
\cite{tube_exchange}. The coupling terms between the rings are
considerably smaller than the inter-ring coupling and therefore can
be neglected in a first approximation. Effectively the system can be
described as a nine site anti-ferromagnetic spin-$\tfrac{1}{2}$
Heisenberg ring showing nearest-neighbor and next-to-nearest
neighbor interactions. The Hamiltonian can be written as \be
H:=\sum_{k=1}^9
\frac{C_1}{4}\vec{\sigma}^{(k)}\cdot\vec{\sigma}^{(k+1)}+\frac{C^{k}_2}{4}\vec{
\sigma}^{(k)}\cdot\vec{\sigma}^{(k+2)}, \label{Hnanotube}\ee with
periodic boundary conditions and approximately homogeneous
parameters for the nearest-neighbor interactions ${C_1=200K}$ and
${C^{k}_2=140K}$ for ${k=2,3,5,6,8,9}$, while ${C^{k}_2=0}$ in all
other cases (see Fig.~\ref{tube_coupl}). The magnetic susceptibility
of this simplified model coincides well with the experimental
results above a temperature of about 10K \cite{tube_exchange}.

For the given Hamiltonian, the thermal state is entangled for low
temperatures and will become separable at a certain point when
increasing the temperature. For every separability criterion, a
critical temperature ${T_c}$ can be found. Doing so for the spin
squeezing inequalities shows that the critical temperature of the
inequalities Eqs.~(\ref{Jxyzineq_spsq3}) is
${T^{(\ref{Jxyzineq_spsq3})}_c=182.8K}$ while the inequality
Eq.~(\ref{Jxyzineq_singlet}) gives
${T^{(\ref{Jxyzineq_singlet})}_c=363.6K}$, the other ones do not
detect any entanglement at all. The critical temperature of
Eq.~(\ref{Jxyzineq_singlet}) has already been known from
Ref.~\cite{tube_ent}, where the magnetic susceptibility of the
system has been used as an entanglement witness, which effectively
results in the same criterion \cite{TK07,njp}. Furthermore, we have
computed the critical temperature of the Peres-Horodecki (PPT)
criterion according to all bipartite splittings, resulting in a
maximal temperature of ${T^{PPT}_c=303.9K}$ for the splitting
$A=\{1,3,4,6,7,9\}$ vs.~$B=\{2,5,8\}.$ So we find a transition from
free to bound entanglement at approximately room temperature.

\section{Conclusions}

We presented a family of entanglement criteria that detect any
entangled state that can be detected based on the first and second
moments of collective angular momenta. We also showed that these
criteria can be extended such that they detect all entangled states
that can be detected based on knowing the expectation values of the
spin components and the correlation matrix. In spite of that these
criteria do not contain multi-qubit correlation terms, they do not
merely detect the entanglement of the two-qubit reduced state. They
can even detect entangled states with separable two-qubit matrix.
For further research, it would be very interesting to extend our
results to ensembles of particles with a higher spin, e.g.
spin-1 particles.

\section*{ACKNOWLEDGMENTS}

 We thank A.~Ac\'{\i}n, A.~Cabello, M.~Christandl, J.I.~Cirac,
S.R.~de Echaniz, K.~Hammerer, S.~Iblisdir, M.~Koschorreck,
J.~Korbicz, J.I.~Latorre, M.~Lewenstein, M.W.~Mitchell,
L.~Tagliacozzo, and R.F.~Werner for fruitful discussions. We also
thank the support of the EU (OLAQUI, SCALA, QICS), the National
Research Fund of Hungary OTKA (Contract No. T049234), and the
Hungarian Academy of Sciences (Bolyai Programme). This work was
further supported by the FWF and the Spanish MEC (Ramon y Cajal
Programme, Consolider-Ingenio 2010 project ''QOIT'').


\begin{thebibliography}{99}
\bibitem{spinsqremark} Note that for Eq.~(\ref{eqKU}) we assume
that the mean spin points to the $x$ direction. Without this requirement,
even a pure state in which all spins point into the same direction could
violate Eq.~(\ref{eqKU}) \cite{K93}.
\bibitem{graphbeispiel}
For example, the reduced states of three neighbored qubits of a
ring cluster state are given by $\vr= (\eins+\sigma_z \sigma_x
\sigma_z)/8,$ which is fully separable. However, the ring cluster
state is the only global state that is compatible with these reduced
states, and is highly entangled.
\end{thebibliography}
\end{document}